\documentclass[12pt,preprint]{aastex}
\usepackage{graphicx}

\shorttitle{IC 4406: a radio-infrared view}
\shortauthors{Cerrigone et al.}

\begin{document}

\title{IC 4406: a radio-infrared view}

\author{Luciano Cerrigone\altaffilmark{1,2} and  Joseph L. Hora\altaffilmark{1}}
\affil{Harvard-Smithsonian Center for Astrophysics, Cambridge, MA 02138 USA} 
\affil{University of Catania, Catania, Italy}

\author{Grazia Umana\altaffilmark{3} and  Corrado Trigilio\altaffilmark{3}}
\affil{INAF, Catania Astrophysical Observatory, Catania, Italy}

\begin{abstract}
IC 4406 is a large (about 100$''$ $\times$ 30$''$) southern bipolar planetary nebula, composed of two elongated lobes extending from a bright central region, where there is evidence for the presence of a large torus of gas and dust.
We show new observations of this source performed with IRAC (Spitzer Space Telescope) and the Australia Telescope
Compact Array. The radio maps show that the flux from the
ionized gas is concentrated in the bright central region and originates in a clumpy structure previously observed in H$\alpha$, while in the infrared images filaments and clumps can be seen in the extended nebular envelope, the central region showing toroidal emission. Modeling of the infrared emission leads to the conclusion that several dust components are present in the nebula. 
\end{abstract}

\section{Introduction}
\label{introduction}
IC 4406 is a well-studied southern planetary nebula. It has been imaged with several telescopes at different wavelength ranges. Near-IR images show two H$_2$ lobes \citep{storey},  orthogonal to the nebula's major axis and  $\sim$25$''$ away from each other. These peaks are approximately coincident with the two blobs observed in 
H$\alpha$+[\ion{N}{2}] and [\ion{O}{3}] \citep{sahai}, interpreted as indicative of the presence of a dense equatorial torus of dust. The optical images show a central ionized region about 32$''$ in diameter. CO maps  show the presence of a collimated high velocity outflow in the polar direction and with  [CO]/[H$_2$]$\approx 5\times 10^{-6}$ and a total molecular mass in the range 0.16--3.2 M$_\odot$ \citep{sahai}. Hubble Space Telescope (HST) WFPC2 images in [\ion{N}{2}], H$\alpha$ and [\ion{O}{3}] have revealed the existence of an intricate system of dark lane features, which led to the name of \lq\lq Retina Nebula\rq\rq~for this object \citep{odell}. The nebula appears to be chemically homogeneous, as \citet{corradi} found no evidence of radial variation for He, O, N, Ne, and Ar. \citet{cox} have detected several C-rich features at mm wavelengths, such as CN, HCO$^+$, HCN and HNC, which indicate the nebula is C-rich, although a C/O ratio of 0.6 is reported by \citet{cohen}.

IC 4406 is a relatively low electron density nebula. Values in the 400-2000 cm$^{-3}$ range have been estimated using several different optical and infrared lines, with values derived by [\ion{S}{2}] and [\ion{O}{3}] doublets matching around 540 cm$^{-3}$ \citep{liu,wang}.
Its central star has a \ion{He}{2} Zanstra temperature of 96800 K \citep{phillips} and its distance is probably around 1.6 kpc \citep{sahai}, although some authors claim it may be overestimated \citep{odell}.

\citet{gruenwald} have modeled IC 4406 with a 3-D photoionization code and fit many observed line intensities assuming there is a torus around the central star. They find as a best fit a central star temperature of $8\times10^4$ K, luminosity of 400 L$_\odot$, torus density 1500 cm$^{-3}$ and nebular density 100 cm$^{-3}$.

In general, comparisons of IR images of planetary nebulae, which trace the molecular gas and warm dust emission, to optical  line images, which trace the ionized gas, have shown the presence of similar structures \citep{latter}, leading to the conclusion that molecular and ionized gas spatially coexist in planetary nebulae, as well as dust grains, despite the different physical conditions these components are presumed to survive in. We have observed IC 4406 in the radio range to inspect the distribution of the ionized gas in its envelope and in the infrared to check for emission from the equatorial dust and molecular gas.

In \S\ref{observations} we explain how we performed our observations and reduced the data; in \S\ref{results} we show our results and in particular in \S\ref{sed} how we modeled the emission in the radio and infrared ranges; \S\ref{nebular} compares our model results to the nebular parameter values obtained directly from the observational data; in \S\ref{summary} we summarize the present work.

\section{Observations and Data reduction}
\label{observations}
\subsection{Radio observations}
Radio observations were performed at the {\bf A}ustralia {\bf T}e\-le\-scope {\bf C}ompact {\bf A}rray (ATCA)\footnote{The Australia Telescope Compact Array is part of the Australia Telescope which is funded by the Commonwealth of Australia for operation as a National Facility managed by CSIRO.} on November 24, 2005 (17:00:00--08:00:00 UT) and December 11, 2005 (15:30:00--02:00:00 UT), simultaneously at 4.8 and 8.6 GHz.
The November run was performed with the array in 1.5C configuration, while for the December one the configuration
was 6.0A. The adopted configurations are both linear but with different antenna positions giving maximum baselines of
4500 m (1.5C) and 5939 m (6.0A), minimum baselines of 77 m (1.5C) and 337 m (6.0A). The pre-calibration of the
array was performed observing 0823-500, while the absolute flux calibrator was 1934-638. Another target was also
observed during our two runs and the total on-target time was about 7 hours for each of the two. The phase calibrator
chosen for IC 4406 was 1431-48, which is 4$^{\circ}$.76 away from the target. The data were reduced with {\bf MIRIAD}, following a
standard reduction procedure as recommended in the MIRIAD User's Guide. The data from the two runs were
combined into one dataset, obtaining a {\it uv} coverage from 0.9 to 96 k$\lambda$ at 4.8 GHz and from 1.5 to 172 k$\lambda$ at 8.6 GHz.
These correspond to an angular resolution of 2$''$.2 at 4.8 GHz and 1$''$.2 at 8.6 GHz, while the largest observable
structures (Largest Angular Scale) are 230$''$ and 140$''$ respectively. Such a setup suits the need to collect all the radiation from the target, whose maximum size, as previously mentioned, is about 100$''$. 

After combining the datasets from the two runs, we created a dirty map from the \textit{uv}-dataset with the task INVERT. The image size was set to 1024 pixel in X band (8.6 GHz) but a larger size (2048 pixel) was necessary for the C band (4.8 GHz), to include a secondary source in the field and allow subsequent proper cleaning of the map. The cell size was set to 0$''$.7 at 4.8 and 0$''$.4 at 8.6 GHz, so that the beam would be extended over $\sim$3 cells. We set the INVERT parameter \textit{options} to \textit{mfs, double}, so that the map could be obtained from a multi-frequency dataset without averaging in frequency and the size of the beam, created along with the map, would be the double of the map size, for a better performance of the cleaning algorithm in the following reduction step. To combine sensitivity to extended emission with sidelobe suppression, we used Brigg's weights, leaving the parameter \textit{sup} unset and \textit{robust}=1. Smaller values of the ROBUST parameter determine in our maps a  worse signal-to-noise ratio. This setting gave a beam of 4$''$.122$\times$3$''$.089 at 4.8 and 2$''$.669$\times$2$''$.060 at 8.6 GHz.

We then used the task CLEAN to deconvolve the dirty map from the synthetic beam. We set a \textit{gain} of 0.1 and \textit{minpatch} 257 for each CLEAN cycle, letting the task choose the proper algorithm, which at both frequencies was the CLARK one. We performed 3000 CLEAN iterations, to reach the theoretical rms noise, as given in the output by INVERT.

As at both frequencies our target appears to be quite resolved, the estimate of its flux has been performed on naturally weighted maps obtained using tapering (task INVERT: fwhm=15, for both frequency bands). This results in larger weights for the visibility points corresponding to short baselines, then limits the chance to miss extended flux in the final map, although possibly producing higher noise. Since the error in our measurement is primarily determined by the error in the absolute flux calibration, tapering is not an issue. With such procedure we measure at 4.8 GHz 103.3 $\pm$ 0.3~mJy and 92.5 $\pm$ 0.4 mJy at 8.6 GHz, which, taking into account a typical 5\% error in the absolute calibration, gives us the following final measurements: 103 $\pm$ 5 mJy at 4.8 GHz and 92 $\pm$ 5 mJy at 8.6 GHz\footnote{The final error has been estimated as $\sigma=\sqrt{rms^2 + (\sigma_{cal}F)^2}$, where $\sigma_{cal}$ is the 5\% relative error in absolute calibration and F is the measured flux density.}. 
These values agree with the previous measurements in \citet{milnealler75}, performed with single dish telescopes, and therefore we can conclude that no flux is missing.

\subsection{Infrared observations}
Infrared observations were performed with the {\bf I}nfra\-{\bf R}ed {\bf A}rray {\bf C}amera (IRAC) \citep{irac} onboard the Spitzer Space Te\-le\-scope (Spitzer) at 3.6, 4.5,
5.8 and 8.0~$\mu$m on March 06, 2004 (UT 09:54:16.311) as part of the GTO program  \lq\lq Studying stellar ejecta on the large scale with SIRTF-IRAC\rq\rq (AOR ID: 4414208). Six High 
Dynamic Range 30 sec dithered frames were obtained
at each wavelength, for a total exposure time of 180 sec per channel. 

Basic Calibrated Data (BCD) were retrieved from Spitzer ar\-chive, cleaned to correct such artifacts as mux-bleeding and
banding and then coadded using {\bf IRACproc} \citep{schuster}.
For flux measurement the images were first converted from MJy/sr (IRAC BCD files are in units of MJy/sr) into Jy units using IRACproc. Then in each image four areas were boxed with a polygon using the task CGCURS in MIRIAD, the mean emission in each polygon was calculated then averaged to obtain an estimate of the background. The standard deviation of the background mean value was assumed as an error. The whole nebula in each image was then boxed with a polygon of approximately $135''\times65''$. The size of the emitting nebula has been determined as the contour at the background+3$\sigma$ level and is approximately: at 3.6 $\mu$m 110$''\times45''$, at 4.5 $\mu$m 120$''\times50''$, at 5.8 $\mu$m 120$''\times60''$ and at 8.0 $\mu$m 125$''\times65''$.  The flux within each $135''\times65''$ polygon was summed up and the same
procedure was adopted for the field star observed West of the central core, so that its flux was subtracted to the overall
nebula's one.  The result was corrected for extended emission according to the IRAC Data Handbook for aperture photometry. The error in the background estimate was assumed as the error of the flux measurement. We have measured at 3.6, 4.5, 5.8, and 8.0 $\mu$m the following fluxes respectively: 64 $\pm$ 2, 101 $\pm$ 4, 122 $\pm$ 7, and 248 $\pm$ 8 mJy.

Table~\ref{tab:fluxes} summarizes the results of our radio and infrared observations.

\subsection{HST archive observations}
\begin{figure}
\centering
\includegraphics[angle=270,scale=0.55]{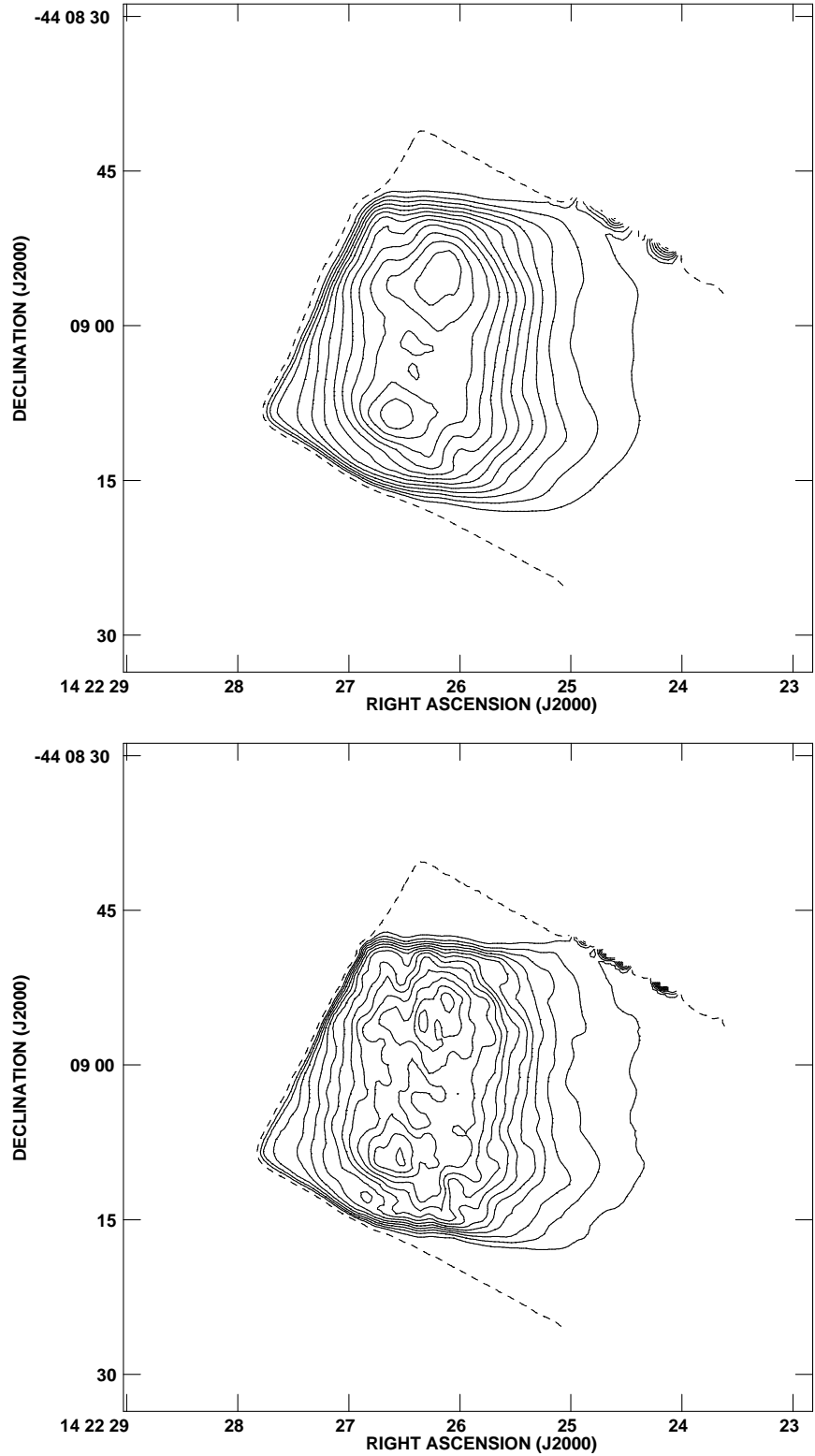}
\includegraphics[angle=270,scale=0.55]{f5b.eps}
\caption{HST WFPC2 image scaled to match our 4.8 (top) and 8,6 GHz (bottom) maps. The units are DN and the absolute values of the levels -3, 3, 6, 9, 12, 15, 18, 21, 24, 27, 30, 33, 36, 39, 42.}
\label{fig:hst_scaled}
\end{figure}

We have retrieved the calibrated H$\alpha$ WFPC2 data from the HST web archive. The only image processing needed was to remove cosmic rays with the IRAF task CRREJ. The WFPC2 array is made of four different arrays, the central one (PC) having a pixel size of 0$''$.0455 and the other three arrays (WF) 0$''$.0996. We selected the image obtained with the PC array and scaled it to match our radio beam (2$''$.669$\times$2$''$.060) with the IRAF task GAUSS (Fig.\ref{fig:hst_scaled}). The astrometry of the WFPC2 image was then modified with the IDL task HASTROM so that it could match our ATCA maps. HST images have a typical position uncertainty of $\sim$0$''$.5 \citep{lee}: we shifted the H$\alpha$ image, to match the radio emission region, of 0$''$.13 E and 0$''$.2 S.

\section{Results and Analysis}
\label{results}
\begin{figure}
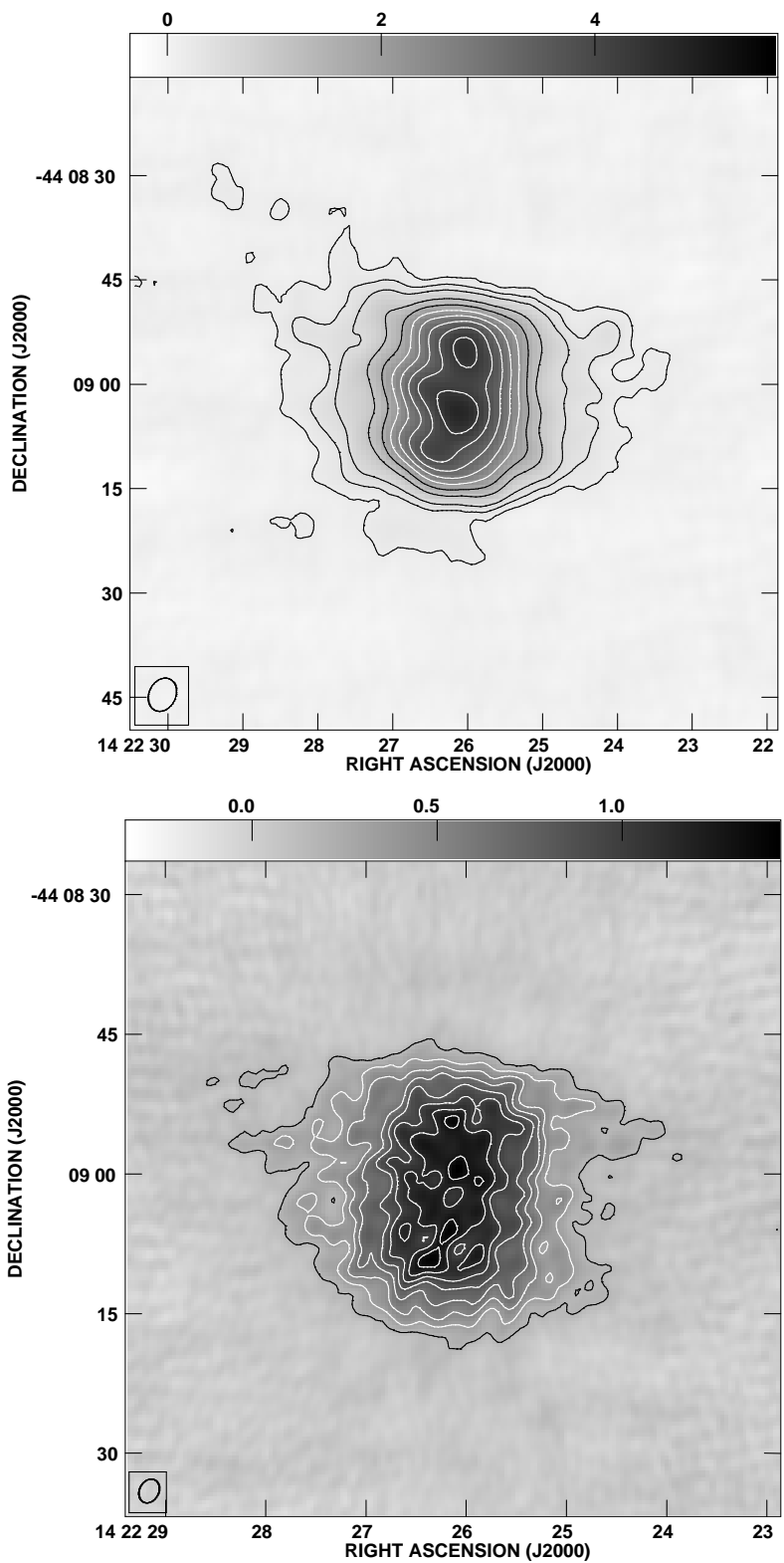

\centering
 \includegraphics[angle=270,scale=0.55]{f1a.eps}
 \includegraphics[angle=270,scale=0.55]{f1b.eps}
\caption{Radio maps of IC 4406 ({\it top}: 4.8 GHz, {\it bottom}: 8.6 GHz). The contours indicate the (-3, 3, 10, 20, 30, 40, 50, 60, 70, 80, 90, 100, 110)$\sigma$ and the (-3, 3, 6, 9, 12, 15, 18, 21, 24, 27)$\sigma$ levels for the top and bottom map respectively, with $\sigma \sim$0.05 mJy for both maps. Both maps were produced with uniform weighting and ROBUST=1. The synthetic beam is shown in the bottom left of each map and the flux density unit is Jy/beam.}
\label{fig:radiomaps}       
\end{figure}

Our radio maps (Fig.\ref{fig:radiomaps}) show the presence of a 42$''$ $\times$ 56$''$ (3$\sigma$ level) emitting region at 4.8 GHz, elongated in E-W direction; at
a 10\% of the peak level the size of the emitting region is restricted to 32$''$ $\times$ 32$''$. At 8.6 GHz the 10\% of the peak level
gives a size of 36$''$ $\times$ 40$''$, in agreement with its 3$\sigma$ level size.
The maps do not show any N-S blobs of emission. What is seen is a clumpy emitting region that resembles what is observed by HST  in H$\alpha$ \citep{odell}.

Using the fluxes that we estimate at the two frequencies, we can calculate a spectral index $\alpha=-0.19\pm0.09$, which matches the expected value of -0.1 for an optically thin radio shell.

\begin{figure}
\centering
\epsscale{0.75}
\plotone{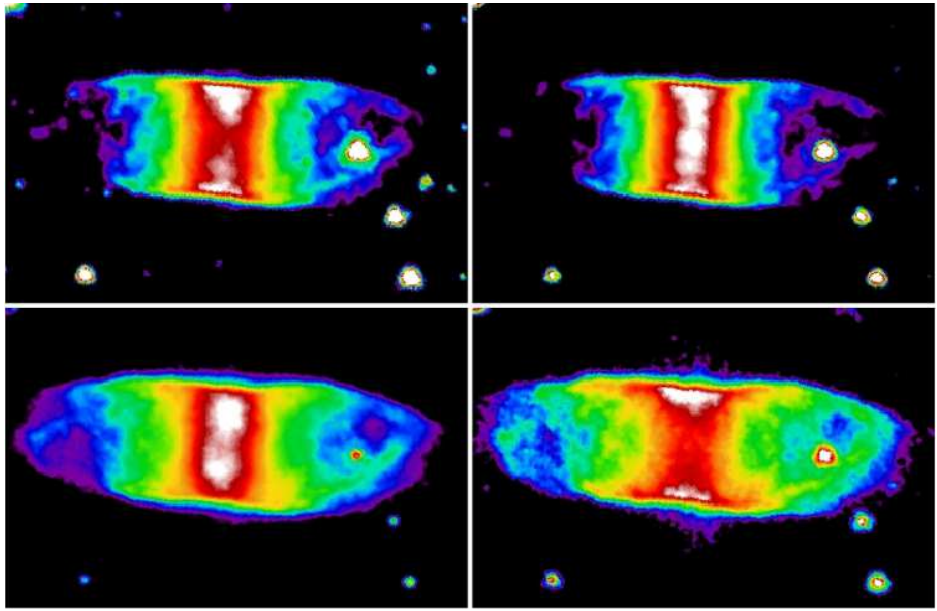}
\caption{
The emission observed with IRAC.
The scale is logarithmic: the maximum (white) is the emission peak in the area,
the minimum (purple)  is 3$\sigma$ (maxima are respectively for 3.6, 4.5, 5.8 and 8.0~$\mu$m: 3.97, 5.85, 6.75 and 15.07 MJy/sr, while $\sigma$ is 0.09, 0.14, 0.14, and 0.52 MJy/sr). Clockwise from top left: 3.6,
4.5, 5.8, 8.0 $\mu$m.}
\label{fig:composite}       
\end{figure}

Fig.\ref{fig:composite} shows the IRAC images of the nebula. To properly view the central equatorial area, the images were plotted with a logarithmic scale having the
peak flux and 3$\sigma$ of the image as thresholds. Channel 1 and especially Channel 3 resemble the H$_2$ emission image in \citet{storey} and this is probably because the $\nu$ = 0--0 S(7) line at 5.51 $\mu$m falls within the 5.8 $\mu$m band and several H$_2$ lines may contribute to the 3.6 $\mu$m image \citep{hora}. Channel 4 clearly shows the emission from the torus of dust surrounding
the central star. Its size is about 28$''$ $\times$ 20$''$, elongated in N-S direction and the angular distance between its peaks is
about 14$''$ (the north peak is found at $14^h22^m26^s.05$, $-44^\circ08'54''.64$ and the south one at $14^h22^m26^s.31$, $-44^\circ09'08''.14$, with a position angle of 168$^\circ$.7). The overall size of the torus matches  the approximate size of the nebula in the N-S direction ($\sim$30$''$), although its peaks are much closer to the center than the H$_2$ blobs reported by \citet{storey}, whose separation can be roughly estimated as $\sim$25$''$. This indicates that the torus is partly shielding the molecular gas from the UV radiation from the central star.

\begin{figure}
\centering
\epsscale{0.5}
\plotone{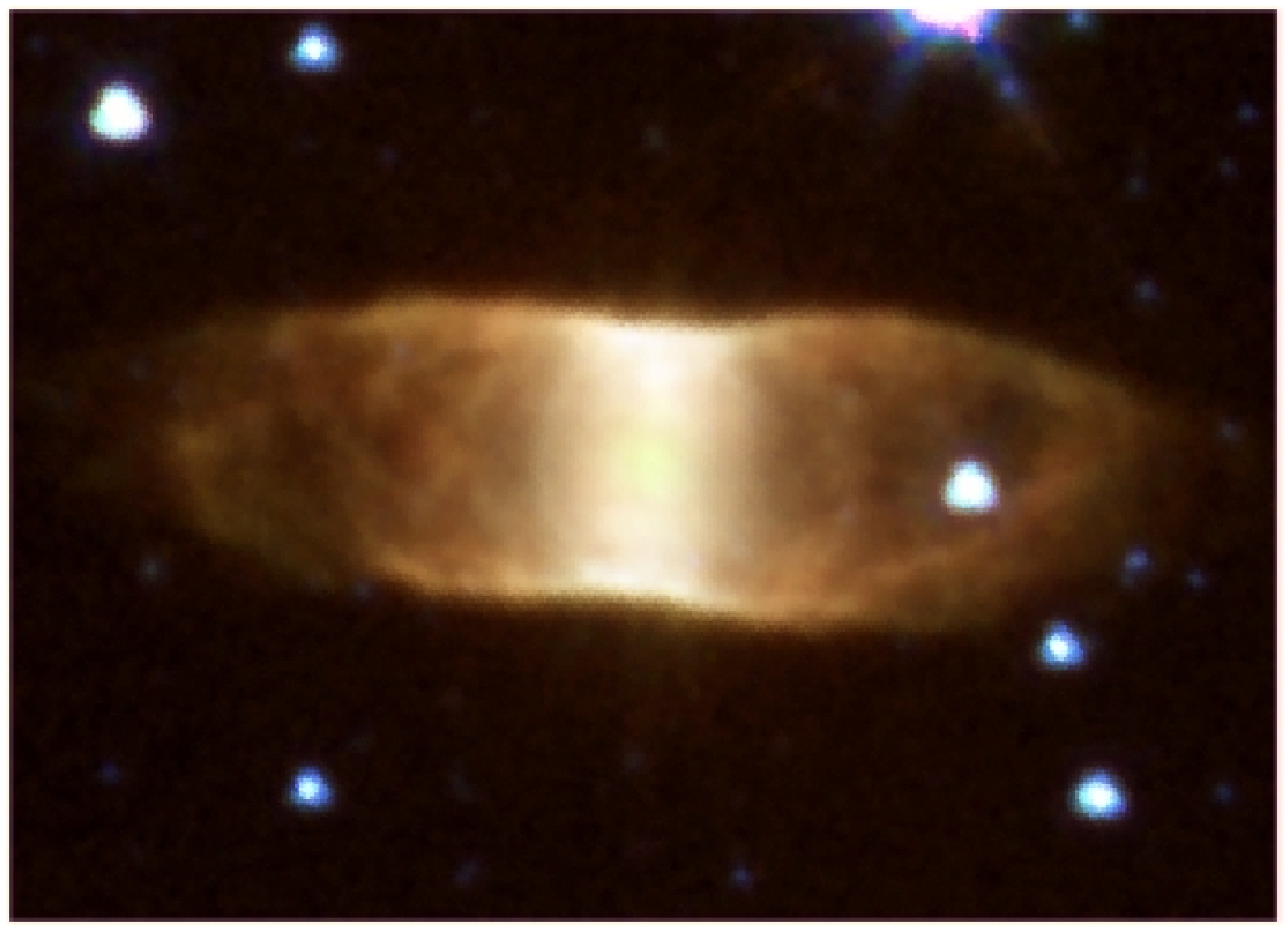}
\plotone{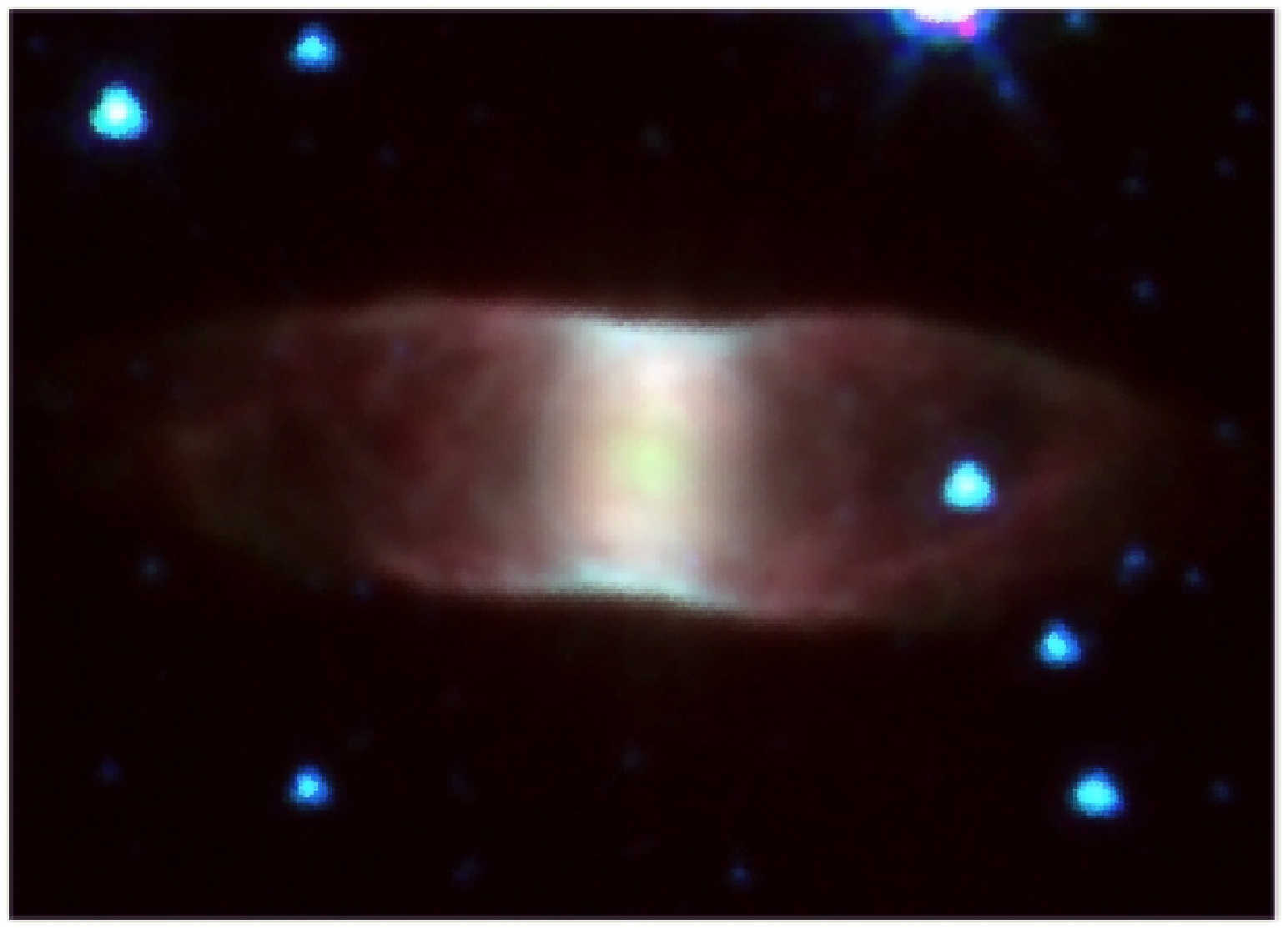}
 \plotone{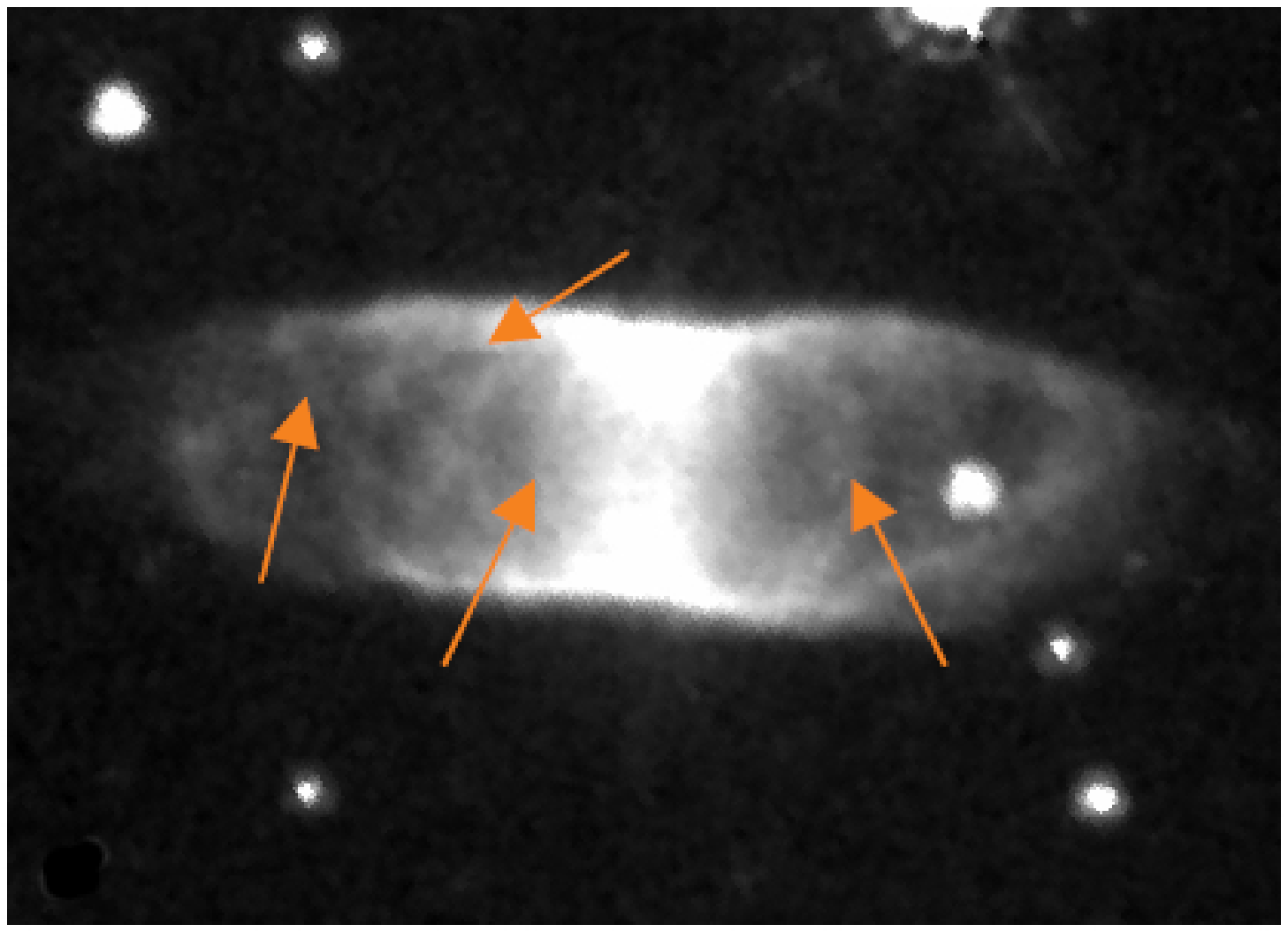}
\caption{Images obtained with IRAC. {\it Top}: overlay of IRAC channels (blue is 3.6, green 4.5, orange 5.8 and red is 8.0 $\mu$m). {\it Middle}: as in the {\it top} image but not including 5.8 $\mu$m channel, for better viewing the central core emission. {\it Bottom}: 5.8 $\mu$m image scaled for a better view of the filament structures, some of which are pointed by the arrows.}
\label{fig:irac}       
\end{figure}

\begin{figure}
\centering
\includegraphics[angle=270,scale=0.65]{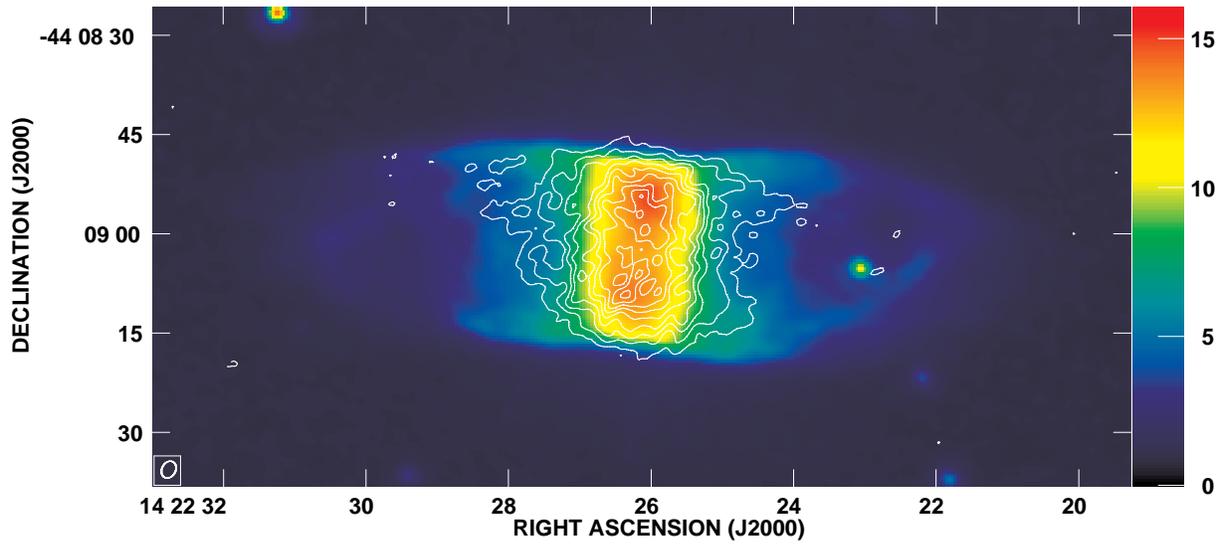}
\caption{IRAC channel 4 image is shown in false color (units are MJy/sr) and overlayed are the 8.6 GHz contours as in Fig.\ref{fig:radiomaps}.}
\label{fig:overlap} 
\end{figure}

Fig.\ref{fig:irac} is a combination of IRAC channels plotted with linear scale. Despite the lower
resolution compared to Hubble images, IRAC is able to detect the faint emission from the neutral components in the
envelope and reveals the structure of the elongated lobes. The IRAC images show filaments at different distances and inclinations
from the central star, connected to the mass loss history of the nebula. The filaments that are closer in projected separation to the central star
show relatively stronger blue (3.6 $\mu$m) emission, which may imply a higher temperature,  being intrinsically
closer to the central object. The overall structure observed in the envelope corresponds to the assumption that the central torus is the main collimating agent, as confirmed by the superposition of the 8.6 GHz and 8 $\mu m$ images in Fig.\ref{fig:overlap}.

\begin{figure}
\centering
\plotone{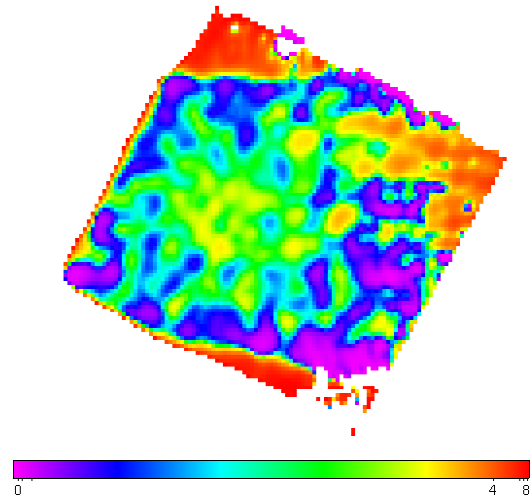}
\caption{Optical depth map of the central region, obtained from the 8.6 GHz map and the H$\alpha$ image following \citet{lee}.}
\label{fig:tau}
\end{figure}

The scaled WFPC2 images match reasonably well our radio maps at 6 and 3 cm.  Following \citet{lee} we also calculated the expected H$\alpha$ image from our 8.6 GHz radio map and then an optical depth map (Fig.\ref{fig:tau}).  The optical depth map confirms the clumpy nature of the central region and it also points out to a larger absorption toward the very core of the region, implying that dust can be present even in \lq\lq close proximity\rq\rq~to the central star.



\subsection{The Spectral Energy Distribution of IC 4406}
\label{sed}
To inspect the dust properties of our target we have collected literature data that, along with our observations, enable us to build the SED.
We have retrieved 2MASS \citep{2mass} images (J, H and Ks) from the 2MASS archive to measure our target's flux in such bands: the values in the Point Source Catalog neglect the emission from the extended envelope.
To estimate the flux and its error a procedure analogous to that used for IRAC data was applied. In each image four areas were boxed with a polygon, then the mean emission in each polygon was calculated, averaged to obtain the background and its standard deviation was taken as the flux density error for the selected image. The nebula itself was boxed with a polygon around its background+3$\sigma$ contour, the flux within the polygon was calculated then background subtracted (the polygon size was the same as for the IRAC data). 2MASS fluxes were then converted into UKIRT system J, H and K magnitudes, so that color correction according to \citet{schlegel} could be performed, adopting E(B-V)=0.19 \citep{gathier}. 
The IRAS color corrected data were taken from \citet{sahai}. In our modeling we have assumed a central star temperature of 96800 K \citep{phillips}, a distance of 1.6 kpc, and a luminosity of 170 L$_\odot$ \citep{sahai}.

We have modeled the SED separately for the radio (ATCA and single dish literature data) and infrared (IRAS, IRAC, 2MASS) emission. The radio data collected from the literature are listed in Table~\ref{tab:radiolit} with their references. For the former we have solved the radiation transfer equation in a spherical shell. The density distribution profile in the shell was determined fitting the density profile found in \citet{corradi}, which gave us a radial variation as $r^{-1.252}$. We introduced this density radial variation in our model shell, and found as a best fit to the data a density at the inner radius of 730 cm$^{-3}$, inner radius 18$''$, outer radius 24$''$, having fixed a distance of 1.6 kpc and electron temperature 10$^4$ K. This gives us an optically thin spectrum down to 800 MHz, which confirms the status of our target as a fairly evolved object, and it also provides us with an estimate of the ionized gas mass of about 0.22 M$_\odot$. To estimate the free-free contribution from radio to near-IR wavelengths, in our model we have calculated the Gaunt factor according to  \citet{gaunt}.

For the infrared range we have used the code DUSTY \citep{dusty} to solve the radiation transfer, assuming once again the nebula to be spherical.
This simple assumption can provide reasonable constraints to the main properties of the envelope, since asymmetries in the density distribution would mostly affect the optical part of the SED, not the mid-IR/FIR region that we are modeling \citep{sanchez}. DUSTY does not allow the simultaneous treatment of more than one shell, yet its output can be used as an input in a second run, thus mimicing the  shell structure of the nebula. 

One constraint to our model is the optical depth at a specified wavelength, which we can calculate as
\begin{equation}
\tau_\nu=2.03 \times 10^{10} \frac{F_\nu}{\theta^2 B_\nu(T_{d})}
\label{eqn:opticaldepth}
\end{equation}
where $T_{d}$ is the dust temperature, $\theta$ the angular radius of the nebula in arcsec, $F_\nu$ the flux density at the frequency $\nu$ in erg cm$^{-2}$ s$^{-1}$ Hz$^{-1}$, $B_\nu(T_{d})$ the Planck function at the temperature $T_{d}$ in erg cm$^{-2}$ s$^{-1}$ Hz$^{-1}$ sr$^{-1}$ \cite[]{gathier86}. 

The first attempts to fit the data were done with a standard MRN \citep{mrn} size distribution of the grains, with $a_{min}=0.005$, $a_{max}=0.26$, $n(a) \propto a^{q}$, $q=-3.5$ , density distribution in the shell as $r^{-2}$, $r$ being the shell radius, a chemical composition of amorphous Carbon only and 0.1 as optical depth at 0.55 $\mu$m. The choice of an am-C only chemistry is due to the detection of several C-rich features mentioned in Section 1.

By a first inspection of the observed data points, it was evident that the data could not be matched by a single dust component. In order to fit several components, we have performed our fit in steps, fitting first the lower wavelength data in a DUSTY run reproducing a hot inner shell, then giving the output of this run as an input to a second run of DUSTY. We have been able to reproduce the observed data assuming the dust envelope is made up of three shells, containing hot, warm and cold dust.
The temperature of the cold component thus obtained was used in Eq.\ref{eqn:opticaldepth} to calculate $\tau$ at 60 $\mu$m. We chose this wavelength because  at this wavelength all the flux seems to be due to one emitting component and cirrus contribution is negligible, which is not necessarily true at 100 $\mu$m. Having calculated the optical depth at 60 $\mu$m with Eq.\ref{eqn:opticaldepth}, we checked if this value matched the one given by DUSTY in its output. This was not the case. Then we used the $\tau_{60}$ estimate as an input optical depth in DUSTY. This led to mismatch all the longer wavelength data points. We started changing the cold dust temperature, looking for a combination of $T_d$ and $\tau_{60}$ that would allow to match the data. We found that it was not possible to reach such a match with the specified set of input parameters, the DUSTY fluxes at larger wavelengths being larger than the observed ones. We have then tried to change the density distribution dependence on the radius: such exponents as -3, -1, -0.5 were tested but none resulted into a good match to the observations. Our second attempt was changing the grain size: we  noticed that the presence of larger grains in the model (up to 6.5 $\mu$m) could effectively modify the reproduced data.

Since DUSTY makes use of spherical geometry, we have assumed an effective radius of 45$''$, corresponding to the radius of a circle having the same area as the dust ellipse observed in our IRAC images. The shell relative thickness parameter in DUSTY has been calculated to reproduce this angular size. The final set of parameters for our best fit is reported in Table~\ref{tab:fit}.

\begin{figure}
\centering
\hspace{-1cm}
\epsscale{0.9}
\plotone{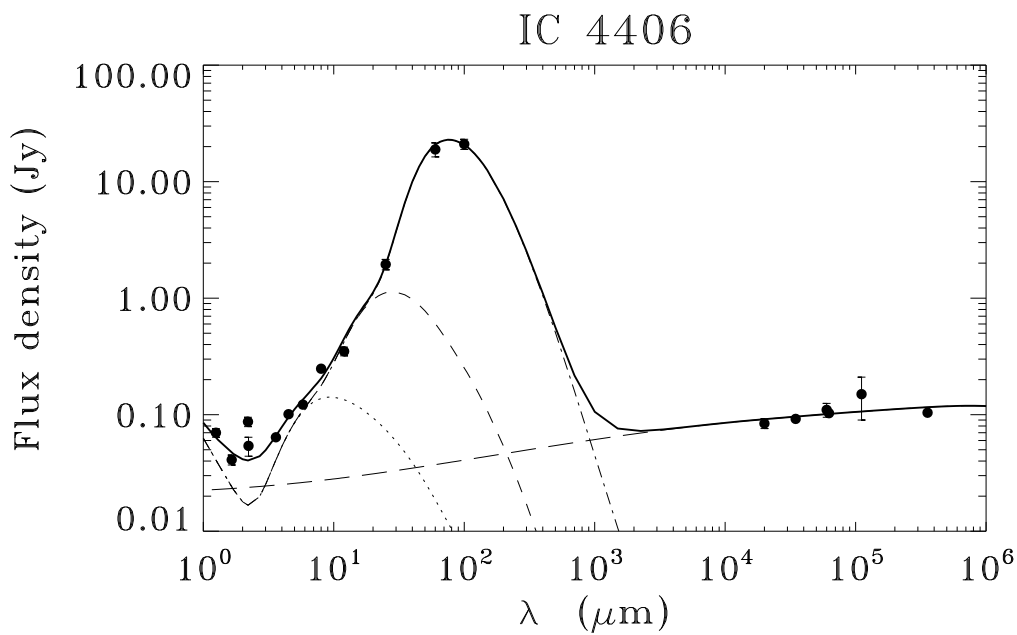}
\caption{SED of IC 4406 from near-IR to radio wavelengths. The best fit is obtained summing the DUSTY and free-free continuum model outputs, and is shown as a thick solid line. The single fit components are also shown. The {\it dot} curve is the result of a model with the  central star and the 700 K component, the {\it short dash} curve includes the central star, 700 K and 200 K components and  the {\it dot-dash} curve includes the central star and the 700, 200, and 57 K dust components, the {\it long dash} curve is the free-free continuum. Error bars are shown for all data, some are smaller than the data point.}
\label{fig:3compzoom}
\end{figure}

Fig.\ref{fig:3compzoom} shows the fit components to the observational data points and the combination of the infrared and radio fits. 

We notice that, if in our free-free model we had used the usual radio approximation of the Gaunt factor, the model would predict around 2 $\mu$m a lower level of emission than observed (as can be seen in Figure~\ref{fig:3compzoom}, where the single emitting components are plotted), which might be interpreted as due to a fourth hotter component of dust missing in the model. Our proper estimate of the Gaunt factor shows how the free-free contribution in the near-IR is actually non negligible and allows us to achieve a good fit to the data points in this range. The fairly larger flux measured in K band can be explained when considering that H$_2$ emission has been detected in IC 4406 \citep{storey} and several lines may fall within the K band filters, along with ionized gas lines such as Br$\gamma$. In fact \citet{phillips05} and \citet{ramoslarios} have explained the excess in Ks band in terms of H$_2$ emission, in particular the latter show how the Ks band 2MASS image matches the H$_2$ image in \citet{storey}.


\subsection{Nebular parameters}
\label{nebular}
Our models enable us to estimate such nebular parameters as electron density, ionized gas mass and dust mass. The dust mass can be calculated from the DUSTY output following \citet{sarkar} as
\begin{equation}
M_d=4\pi R^2Y\frac{\tau_{100}}{k_{100}}
\label{modeldust}
\end{equation}
where R is the inner radius of the emitting shell in cm, Y is the thickness of the shell relative to R, $\tau_{100}$ and k$_{100}$ the optical depth and absorption coefficient at 100 $\mu$m. We can now use our DUSTY output for R and $\tau_{100}$, which, for the 57 K more external shell, are estimated as 3.6$\times10^{16}$ cm and 2.34$\times10^{-5}$; Y is 30 and k$_{100}$=92 cm$^2$ g$^{-1}$, calculated following \cite{jura}. We thus obtain a dust mass of 6$\times$10$^{-5}$ M$_\odot$. Our radio model gives instead an average electron density of 450 cm$^{-3}$ and ionized mass of 0.29 M$_\odot$, as mentioned in the previous section.

These values can be compared to those derived by equations that directly use the observed fluxes. 
From our radio observations it is possible to derive the H$_\beta$ flux and electron density, which can then be used to estimate the ionized mass of the nebula.

\begin{equation}
H_\beta=\frac{S_{4.8\,GHz}}{2.83\times10^9 \, t^{0.53} [1+(1-x)y+3.7xy]} \;\; erg\, cm^{-2} s^{-1}
\label{eqn:hbeta}
\end{equation}

\begin{equation}
n_e=2.74\times 10^4\sqrt{\frac{t^{0.88} H_\beta}{\epsilon \theta^3 d}} \;\; cm^{-3}
\label{eqn:density}
\end{equation}

\begin{equation}
M_{ion}=\frac{11.06 \,d^2 t^{0.88}}{n_e} H_\beta \;\; M_\odot
\label{eqn:ionmass}
\end{equation}

In Eq.\ref{eqn:ionmass} H$_\beta$ is the H$_\beta$ line flux in units of $10^{-11}$ erg cm$^{-2}$ s$^{-1}$, n$_e$ the electron density in cm$^{-3}$, $t$ the electron gas temperature in units of $10^4$ K, $d$ the distance to the star in kpc. In Eq.\ref{eqn:density} H$_\beta$ is  like in the previous equation, $\theta$ is the ionized gas radius as deduced from the 4.8 GHz radio map (we have used a value of 24$''$) in arcsec, $\epsilon$ is the filling factor, for which we have used an average value of 0.6, $t$ as before.  In Eq.\ref{eqn:hbeta}, S$_{4.8GHz}$ is the 4.8 GHz flux density in Jy, $t$ as before, x is $He^{++}/He=0.121$, y is $He/H=0.132$, calculated from the abundances in \citet{corradi}. Eq.\ref{eqn:hbeta}, \ref{eqn:ionmass} and \ref{eqn:density} are from \citet{pott1}.
We find 0.21 M$_\odot$, 418 cm$^{-3}$ and 3.09$\times$10$^{-11}$ erg cm$^{-2}$ s$^{-1}$ for the ionized mass, electron density and H$_\beta$ flux respectively, which closely match our radio model's results.

The dust mass can be estimated according to \cite{pott2} as
\begin{equation}
M_{dust}=\frac{4}{3}\frac{a\rho}{Q_{\nu}}\frac{d^2 F_{\nu}}{B(\nu,T_{dust})} \;\; g
\label{eqn:dustmass}
\end{equation}

Following \citet{jura}, we consider 1.1 as a representative value of the power-law distribution of the emissivity of carbon grains in the infrared. We can thus calculate the dust emissivity at 60 $\mu$m assuming average grain radius $a=10^{-5}$ cm and density $\rho=3$ g~cm$^{-3}$, which results into $Q_{60}=1.107 \times 10^{-3}$. Considering the flux at 60 $\mu$m as due only to the cold component at 57 K in our best fit, we calculate a dust mass of about $2.8 \times 10^{-4}$ M$_\odot$. We also calculated the dust mass of the other emitting components in our fit, but they resulted to be negligible when compared to the cold dust: we find for the 700 K component $5 \times 10^{-11}$ M$_\odot$ (calculated using the 4.5 $\mu$m flux) and for the 200 K component $4 \times 10^{-8}$ M$_\odot$ (using the 12 $\mu$m flux). The dust to gas mass ratio can be estimated as $M_d/M_g=1.3 \times 10^{-3}$. See Table~\ref{tab:results} for a summary of the parameters we have derived.

We notice that the cold dust mass value we estimate by DUSTY is one order of magnitude smaller than that calculated with Eq.\ref{eqn:dustmass}.
This could perhaps be due to the approximations intrinsic to both the DUSTY modeling and the derivation of Eq.\ref{eqn:dustmass} (i.e., spherical shape, single component chemistry, physical knowledge of dust opacity). For example, \citet{sarkar} noticed how in general the DUSTY SEDs are not very sensitive to cooler dust at large radii, as demonstrated by the fact that large differences in shell relative thickness values do not determine drastically different SEDs. Therefore it is possible that the code itself is underestimating the amount of dust.

Another issue concerning the use of DUSTY is the assumed spherical symmetry. As previously mentioned, one-dimensional models have been utilized in the literature for
modeling the mid- and far-IR SEDs in planetary nebulae, since asymmetries
in the distribution and orientation have less of an effect than in the
optical part of the SED \citep{sanchez}.

Using more realistic geometries with other codes would
be expected to provide more accurate results, but assumptions must be usually made
on the 3D geometry. The advantages of a better match to the geometry of the
nebula given by these codes can be heavily affected by such assumptions, so
a careful exploration of parameter space consistent with the observations
must be performed in order to guide and interpret the modeling results.

The use of one-dimensional codes as DUSTY is therefore still valid in determining the mean properties of an envelope, within the intrinsic errors of any modeling process.

\section{Summary}
\label{summary}
We have observed IC 4406 in the cm and 3--10 micron ranges. Our radio observations have confirmed the presence of the complicated maze of lanes already observed in H$\alpha$ in the central region of the nebula and have enabled us to calculate several nebular parameters, whose values match the classification for this target as an evolved planetary nebula, in particular its low dust to gas mass ratio and density. IRAC imaging has revealed the presence of filaments in the nebula that were not detected in previous observations.

Our IRAC measurements, combined with literature data at longer and shorter wavelengths, have enabled us to study the SED of the PN IC 4406 and reproduce it with DUSTY. This has revealed that three different dust components are needed to model the data, with temperatures ranging from 57 to 700 K. It has also been necessary to include in the model slightly larger grains than in the standard MRN composition (up to 6.5 $\mu$m) to account for the calculated 60 $\mu$m optical depth. The main limits of our modeled curve are the spherical geometry assumed in DUSTY and the lack of data in the mm and sub-mm ranges, which would give a constraint on the slope of the curve. As we have observed during our trials with DUSTY, the slope of the SED in the sub-mm range changes with the maximum size of the grains included in the model. Unfortunately, in this range observations are available only for a few stars so far: none for our target. 

We can speculate that in such a diversified dust environment, as we find in IC 4406, further lower temperature components may exist and future high sensitivity, high angular resolution observations will give a fundamental contribution to understand the physics of circumstellar envelopes in planetary nebulae.

\acknowledgements
L. Cerrigone acknowledges funding from the Smithsonian Astrophysical Observatory through the SAO Predoctoral Fellowship Program. This work is based in part on observations made with the Spitzer Space Telescope, operated by Jet Propulsion Laboratory under NASA contract 1407. This publication makes use of data products from the Two Micron All Sky Survey, which is a joint project of the University of Massachusetts and the Infrared Processing and Analysis Center/California Institute of Technology, funded by the National Aeronautics and Space Administration and the National Science Foundation. The authors wish to thank the anonymous referee for his/her helpful review. \\
 \\

{\it Facilities:} \facility{ATCA}, \facility{Spitzer (IRAC)}

%
%


\begin{table}
\centering
\caption{IC 4406 measured fluxes.\label{tab:fluxes}}
\begin{tabular}{lcccc}
\tableline
\tableline
{\bf Spitzer} & {\it 3.6} $\mu m$ & {\it 4.5} $\mu m$ & {\it 5.8} $\mu m$ & {\it 8.0} $\mu m$\\
          &    (mJy)       &    (mJy)       &     (mJy)      &      (mJy)   \\
IC 4406 &   64 $\pm$ 2 & 101 $\pm$ 4 & 122 $\pm$ 7 & 248 $\pm$ 8 \\
\tableline
 {\bf ATCA} & & {\it 4.8 GHz} & {\it 8.6 GHz}  \\
 &           &    (mJy)       &     (mJy)      &         \\
IC 4406 & & 103.3 $\pm$ 0.3 & 92.5 $\pm$ 0.4 \\
1431-48 & &  1020 $\pm$ 1 & 650 $\pm$ 1 \\
\tableline
\end{tabular}
\end{table}

\begin{table}
\centering
\caption{IC 4406 radio fluxes from the literature.\label{tab:radiolit}}
\begin{tabular}{lc}
\tableline
\tableline
$\nu$ (GHz) & S$_\nu$ (mJy) \\
\tableline
0.843 & 104 $\pm$ 3 \tablenotemark{a} \\
2.7 & 150 $\pm$ 60 \tablenotemark{b} \\
5 & 110 $\pm$ 15  \tablenotemark{c} \\
14.7 & 84 $\pm$ 8 \tablenotemark{d}   \\
\tableline
\end{tabular}
\tablenotetext{a}{\citet{mauch}}
\tablenotetext{b}{\citet{milnewebster}}
\tablenotetext{c}{\citet{milnealler75}}
\tablenotetext{d}{\citet{milnealler82}}
\end{table}

\begin{table}
\centering 
\caption{DUSTY fit parameters.\label{tab:fit} }
\begin{tabular}{lc}
\tableline
\tableline
\textit{Parameters} & \textit{Values} \\
\tableline
Chemistry & 100\% amorphous Carbon \\
Central source & Blackbody at 96800 K \\
Density distribution & $\propto r^{-2}$ \\
Grain size distribution & $\propto r^{-3.5}$, $a_{min}=0.005 \, \mu$m, $a_{max}=6.5 \, \mu$m \\
Hot component & $T_d=700$ K, $\tau_{60}=4.5 \times 10^{-6}$, $R_{in}=7.17\times10^{13}$ cm, Y=14790 \\
Warm component & $T_d=200$ K, $\tau_{60}=1.0 \times 10^{-5}$,  $R_{in}=1.40\times10^{15}$ cm, Y=750\\
Cold component & $T_d=57$ K, $\tau_{60}=6.8 \times 10^{-5}$, $R_{in}=3.62\times10^{16}$ cm, Y=30 \\
\tableline
\end{tabular}
\tablecomments{Y is the shell thickness relative to the inner radius, which is determined by the code as the layer having temperature $T_d$; $R_{in}$ is the distance to the central star of the layer having temperature $T_d$; $\tau_{60}$ is the optical depth at 60 $\mu$m.}

\end{table}

\begin{table}
\centering
\caption{Calculated nebular parameters.
\label{tab:results}}
\begin{tabular}{lcc}
\tableline
\tableline
 &   Empirical & Models \\
M$_{ion}$ (M$_\odot$) & 0.21 & 0.22 \\
M$_{dust}$  (10$^{-4}$ M$_\odot$) & 2.8 & 0.6 \\
M$_{dust}$/M$_{ion}$ ($10^{-4}$)& 13 & 3 \\ 
Gas density (cm$^{-3}$) & 418 & 730 \\
\noalign{\smallskip}\tableline
\end{tabular}
\tablecomments{ As \lq\lq Empirical\rq\rq~we mean values calculated with the equations in \S\ref{nebular} and with \lq\lq Models\rq\rq~those deduced from the modeling explained in \S\ref{sed}. The model gas density is the value at the inner radius.}
\end{table}

\clearpage
%
%

\clearpage
\begin{thebibliography}{}
\bibitem[Cohen \& Barlow, 2005]{cohen}
Cohen, M. and Barlow, M. J., 2005, \mnras, 362, 1199
\bibitem[Corradi et al., 1997]{corradi}
Corradi, R. L. M, Perinotto, M., Schwarz, H. E. and Claeskens, J.-F., 1997, \aap,  322, 975
\bibitem[Cox et al., 1992]{cox}
Cox, P., Omont, A., Huggins, P. J., Bachiller, R. and Forveille, T., 1992, \aap, 266, 420
\bibitem[Fazio et al., 2004]{irac}
Fazio, G. et al., 2004, \apjs, 154(1), 10
\bibitem[Gathier et al., 1986]{gathier86}
Gathier, G. A., Pottasch, S. R. and Pel, J. W., 1986, \aap, 157, 171 
\bibitem[Gathier \& Pottasch, 1988]{gathier}
Gathier and Pottasch, S., 1988, \aap, 197, 226
\bibitem[Gruenwald et al., 1997]{gruenwald}
Gruenwald, R., Viegas, S. M. and Brogui{\`e}re, D., 1997, \apj, 480, 283
\bibitem[Hora et a., 2004]{hora}
Hora, J. L., Latter, W. B., Allen, L. E., Marengo, M., Deutsch, L. K. and Pipher, J. L., 2004, \apj, 154, 296
\bibitem[Ivezi{\'c} et al., 1999]{dusty}
Ivezi{\'c}, \v{Z}., Nenkova, M. and Elitzur, M., 1999, User Manual for DUSTY, Internal Report, Univ. of Kentucky, accessible at 
{\it http://www.pa.uky.edu/$\sim$moshe/dusty}
\bibitem[Jura, 1986]{jura}
Jura, M., 1986, \apj, 303, 327
\bibitem[Karzas \& Latter, 1961]{gaunt}
Karzas, W. J. and Latter, R., 1961, \apj, 6, 167
\bibitem[Latter et al., 1995]{latter}
Latter, W. B., Kelly, D. M., Hora, J. L. and Deutsch, L. K., 1995, \apjs, 100, 159
\bibitem[Lee \& Kwok, 2005]{lee}
Lee, T.-H. \& Kwok, S., 2005, \apj, 632, 340
\bibitem[Liu et al., 2001]{liu}
Liu, X.-W., Barlow, M.~J., Cohen, M., Danziger, I.~J., Luo, S.-G., Baluteau, J.~P., Cox, P., Emery, R.~J., Lim, T. and P{\'e}quignot, D., 2001, \mnras, 323, 343
\bibitem[Mathis et al., 1977]{mrn}
Mathis, J. S., Rumpl, W. and Nordsieck, K. H., 1977, \apj, 217, 425
\bibitem[Mauch et al., 2003]{mauch}
Mauch, T., Murphy, T., Buttery, H. J., Curran, J., Hunstead, R. W., Piestrzynski, B., Robertson, J. G. and Sadler, E. M., 2003, \mnras, 342(4), 1117
\bibitem[Milne \& Aller, 1975]{milnealler75}
Milne, D. K. and Aller, L. H., 1975, \aap, 38, 183
\bibitem[Milne \& Aller, 1982]{milnealler82}
Milne, D. K. and Aller, L. H., 1982, \aaps, 50, 209
\bibitem[Milne \& Webster, 1979]{milnewebster}
Milne, D. K. and Webster, B. L., 1979, \aaps, 36, 179
\bibitem[O'Dell et al., 2002]{odell}
O'Dell, C. R., Balick, B., Hajian, A. R., Henney, W. J. and Burkert, A., 2002, \aj, 123, 3329 
\bibitem[Phillips, 2003]{phillips}
Phillips, J. P., 2003, \mnras, 344(2), 501
\bibitem[Phillips \& Ramos-Larios, 2005]{phillips05}
Phillips, J. P. and Ramos-Larios, G., 2005, \mnras, 364(3), 849
\bibitem[Pottasch, 1984]{pott1}
Pottasch, S., 1984, Planetary Nebulae - A Study of Late Stages of Stellar Evolution, D. Reidel Publishing Co., Dordrecht
\bibitem[Pottasch et al., 1984]{pott2}
Pottasch, S. R., Baud, B., Beintema, D., Emerson, J., Habing, H. J., Harris, S., Houck, J., Jennings, R. and Marsden, P.,  1984, \aap, 138, 10
\bibitem[Ramos-Larios et al., 2006]{ramoslarios}
Ramos-Larios, G., Kemp, S. N. and Phillips, J. P., 2006, \rmxaa, 42, 131
\bibitem[Sahai et al., 1991]{sahai}
Sahai, R., Wooten, A., Schwarz, H. E. and Clegg, R. E. S., 1991, \aap, 251, 560
\bibitem[Sarkar \& Sahai, 2006]{sarkar}
Sarkar, G. and Sahai, R., 2006, \apj, 644, 1171
\bibitem[S{\'a}nchez Contreras et al., 2007]{sanchez}
S{\'a}nchez Contreras, C., Le Mignant, D., Sahai, R., Gil de Paz, A. and Morris, M., 2007, \apj, 656, 1150
\bibitem[Schlegel et al., 1998]{schlegel}
Schlegel, D., Finkbeiner, D. P. and Davis, M., 1998, \apj, 500, 525
\bibitem[Schuster et al., 2006]{schuster}
Schuster, M. T., Marengo, M. and Patten, B., 2006, SPIE, 6270, 65 
\bibitem[Skrutskie et al, 2006]{2mass}
Skrutskie, M. F., et al., 2006, \aj, 131, 1163
\bibitem[Storey, 1984]{storey}
Storey, J. W. V., 1984, \mnras, 206, 521
\bibitem[Wang et al., 2004]{wang}
Wang, W., Liu, X.-W., Zhang, Y. and Barlow, M. J., 2004, \aap, 427, 873
\end{thebibliography}
\end{document}